\documentclass[%
 reprint,
 superscriptaddress,
 amsmath,amssymb,
 aps,prl,
floatfix,
]{revtex4-1}

\usepackage{graphicx}
\usepackage{dcolumn}
\usepackage{bm}
\usepackage{hyperref}

\begin{document}

\preprint{APS/123-QED}


\title{First Results from Axion Haloscope at CAPP around $10.7 \,\mu\rm{eV}$
}

\author{Ohjoon Kwon} \affiliation{Center for Axion and Precision Physics Research (CAPP), IBS, Daejeon 34051, Republic of Korea}

\author{Doyu Lee} \thanks {Present address: Samsung Electronics, Gyeonggi-do 16677, Republic of Korea}
 \affiliation{Center for Axion and Precision Physics Research (CAPP), IBS, Daejeon 34051, Republic of Korea} 

\author{Woohyun Chung}
 \email{Corresoponding author. \\ gnuhcw@ibs.re.kr}
\affiliation{Center for Axion and Precision Physics Research (CAPP), IBS, Daejeon 34051, Republic of Korea}

\author{Danho Ahn} \affiliation{Department of Physics, KAIST, Daejeon 34141, Republic of Korea} \affiliation{Center for Axion and Precision Physics Research (CAPP), IBS, Daejeon 34051, Republic of Korea}

\author{HeeSu Byun} \affiliation{Center for Axion and Precision Physics Research (CAPP), IBS, Daejeon 34051, Republic of Korea}

\author{Fritz Caspers} \affiliation{CERN, European Organization for Nuclear Research, CH-1211 Genève 23, Switzerland}
\affiliation{ESI (European Scientific Institute) Archamps Technople, F-74160, France}

\author{Hyoungsoon Choi} \affiliation{Department of Physics, KAIST, Daejeon 34141, Republic of Korea}

\author{Jihoon Choi} \thanks{Present address: Korea Astronomy and Space Science Institute, Daejeon, 34055, Republic of Korea} \affiliation{Center for Axion and Precision Physics Research (CAPP), IBS, Daejeon 34051, Republic of Korea}

\author{Yonuk Chung} \thanks{Present address: Department of Nano Engineering, Sungkyunkwan University, Suwon 16419, Republic of Korea}
 \affiliation{Korea Research Institute of Standards and Science, Daejeon 34113, Republic of Korea}

\author{Hoyong Jeong} \affiliation{Department of Physics, Korea University, Seoul 02841, Republic of Korea}

\author{Junu Jeong} \affiliation{Department of Physics, KAIST, Daejeon 34141, Republic of Korea}
\affiliation{Center for Axion and Precision Physics Research (CAPP), IBS, Daejeon 34051, Republic of Korea}

\author{Jihn E Kim}\affiliation{Department of Physics, Kyung Hee University, Seoul 02447, South Korea}

\author{Jinsu Kim} \affiliation{Department of Physics, KAIST, Daejeon 34141, Republic of Korea} \affiliation{Center for Axion and Precision Physics Research (CAPP), IBS, Daejeon 34051, Republic of Korea}

\author{\c{C}a\u{g}lar Kutlu} \affiliation{Department of Physics, KAIST, Daejeon 34141, Republic of Korea} \affiliation{Center for Axion and Precision Physics Research (CAPP), IBS, Daejeon 34051, Republic of Korea}

\author{Jihnhwan Lee} \affiliation{Center for Artificial Low Dimensional Electronic Systems, IBS, Pohang 37673, Republic of Korea}

\author{MyeongJae Lee}\affiliation{Center for Axion and Precision Physics Research (CAPP), IBS, Daejeon 34051, Republic of Korea}

\author{Soohyung Lee}\affiliation{Center for Axion and Precision Physics Research (CAPP), IBS, Daejeon 34051, Republic of Korea}

\author{Andrei Matlashov} \affiliation{Center for Axion and Precision Physics Research (CAPP), IBS, Daejeon 34051, Republic of Korea}

\author{Seonjeong Oh}\affiliation{Center for Axion and Precision Physics Research (CAPP), IBS, Daejeon 34051, Republic of Korea}

\author{Seongtae Park}\affiliation{Center for Axion and Precision Physics Research (CAPP), IBS, Daejeon 34051, Republic of Korea}

\author{Sergey Uchaikin}\affiliation{Center for Axion and Precision Physics Research (CAPP), IBS, Daejeon 34051, Republic of Korea}

\author{SungWoo Youn}\affiliation{Center for Axion and Precision Physics Research (CAPP), IBS, Daejeon 34051, Republic of Korea}

\author{Yannis K. Semertzidis} \affiliation{Center for Axion and Precision Physics Research (CAPP), IBS, Daejeon 34051, Republic of Korea} \affiliation {Department of Physics, KAIST, Daejeon 34141, Republic of Korea}

\date{\today}


\begin{abstract}
The Center for Axion and Precision Physics Research at the Institute for Basic Science is searching for axion dark matter using ultralow temperature microwave resonators. 
We report the exclusion of the axion mass range 10.7126$-$10.7186 $\mu$eV with near Kim-Shifman-Vainshtein-Zakharov (KSVZ) coupling sensitivity and the range 10.16$-$11.37 $\mu$eV with about 9 times larger coupling at 90$\%$ confidence level. 
This is the first axion search result in these ranges. 
It is also the first with a resonator physical temperature of less than 40 mK. 
\end{abstract}
\maketitle
\begin{figure}[t]
\includegraphics[width=\linewidth]{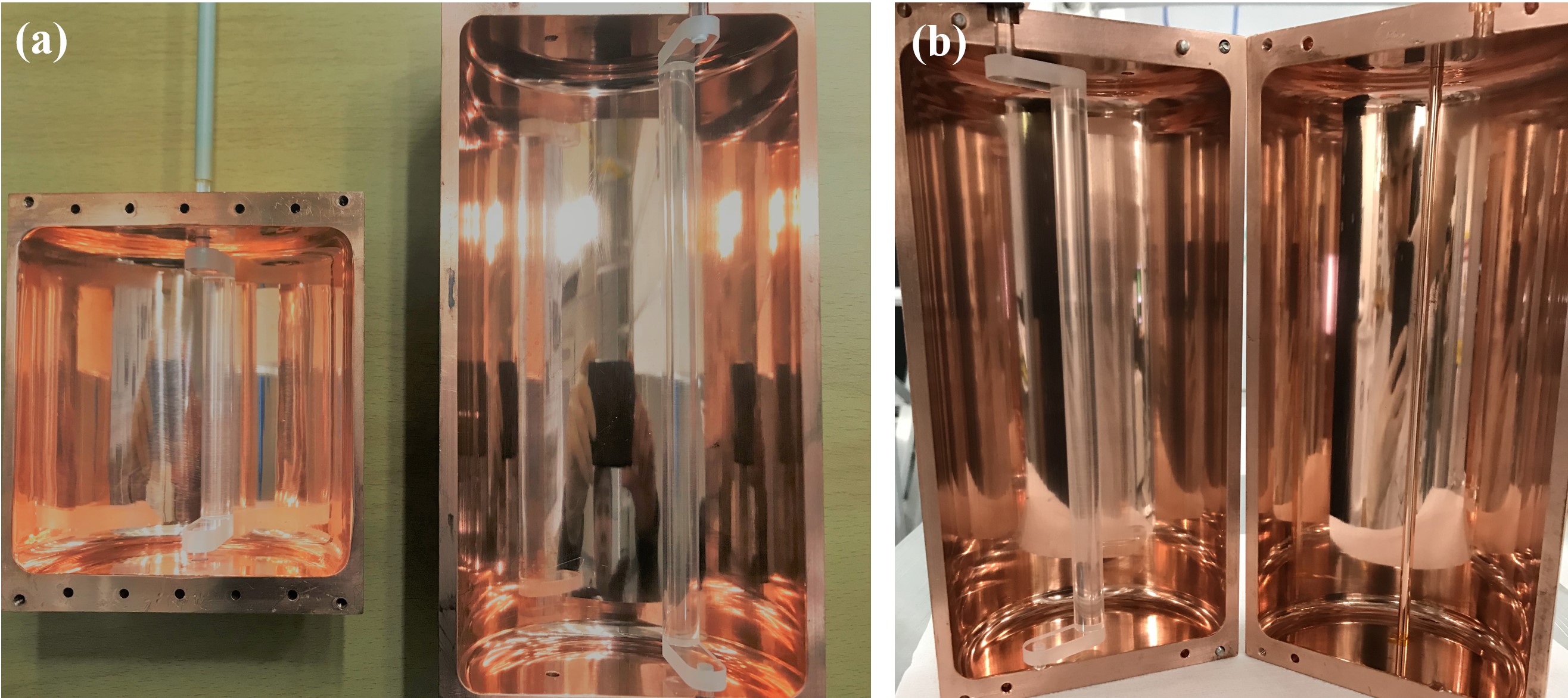}
\caption{Split cavity structures. (a) One half of the 0.59\,L cavity used in 9KSVZ-1 run and a half of the 1.12\,L cavity used in the other runs. (b) The cavity set for the 9KSVZ second run. The sapphire rod is used for frequency tuning. The fixed copper rod for a slight frequency increase is also shown in the right half. }
\label{fig:split_cav}
\end{figure}

%
%
The absence of \textit{CP} violation in strong interactions requires a tiny coefficient of order $\le 10^{-10}$ for the terms contributing to the electric dipole moments of the neutron \cite{rev_par_phys}. 
However, the modern theory of strong interactions, quantum chromodynamics (QCD), generically introduces a coefficient of order one for the \textit{CP} violating vacuum angle $\theta_\text{QCD}$.
The ``\textit{CP} violation" (in addition to \textit{C} violation) of Sakharov \cite{Sakharov:1967dj} necessary to generate the baryon asymmetry of the Universe does not belong to the strong interactions. 
In QCD, Peccei and Quinn (PQ) provided a solution of this strong \textit{CP} problem, by introducing a global $U(1)_\text{PQ}$ symmetry, first making the vacuum angle term irrelevant \cite{PQ}. 
However, we know that all global symmetries should be broken, and for the PQ symmetry the magnitude of breaking at the electroweak scale is such that the mass of the resulting pseudoscalar, called QCD axion, is in the keV range \cite{Wein,Wilc}, which was excluded after Refs. \cite{RevModPhys_2003_Bradley,axion_bound}. \\
\indent 
It has been proposed \cite{KSVZ1979} that axions below a few meV would be long-lived and the interactions being weak enough to be the dark matter of the Universe. 
Two classes in this kind of models, contributing the dark matter in the Universe, are the  Kim-Shifman-Vainstein-Zakharov (KSVZ) \cite{KSVZ1979,KSVZ1980} and the Dine-Fischler-Srednicki-Zhitnitsky (DFSZ) \cite{DFSZ_Zhitnitsky1980,*DFSZ1981} models. 
Currently, several experimental groups are actively working to search for axions \cite{Invisible_axion_search_method_sikivie, Kim_GP_rev_mod}, at the level that axions constitute 100$\%$ of the local dark matter density. 
The most advanced method for over 30 years uses the so called Sikivie haloscope \cite{Sikivie}, searching for cosmic axions converting to photons inside a high-quality resonator immersed in a strong magnetic field. 
Since the first pioneering experiments with this method \cite{RBF1,*RBF2,UF}, there have been numerous cavity experiments in search of axions \cite{ADMX_PRL_1998,*PRD_ADMX2001,*Asztalos_2002,*PRD_ADMX2004,*ADMX_sidecar,*ADMX_PRL_DFSZ,*ADMX_DFSZ_2020,haystac_PRL,Quax_agam,capp8tb, *multicell_prl}. 
However, most of the candidate mass range remains unexplored. \\
\indent 
The Center for Axion and Precision Physics Research (CAPP) uses this cavity method to search for axion dark matter, and we report its Pilot Axion-Cavity Experiment (CAPP-PACE), equipped with an 8 T superconducting magnet, pursuing a low-noise axion haloscope. 
CAPP-PACE aims to apply leading-edge technologies to axion experiments through R$\&$D \cite{capp_jpa,ybco_1}, near and above 10 $\mu$eV.
In this Letter, limits are given for the axion masses near 10.7 $\mu$eV. We minimized the physical temperature of the cavity down to 38 mK. This is the coldest axion dark matter experiment to date. 
We obtained the result with a high electron mobility transistor (HEMT) amplifier with a noise temperature around 1 K. 
The subsequent use of a near quantum
noise limited Josephson parametric amplifier (JPA), employing the latest R\&D results of CAPP \cite{sergey_jpa, capp_jpa}, improves the scanning speed by more than an order of magnitude in the next phase of the experiment.  \\%
%
%
%
%
%
%
\indent 
When axions couple with a magnetic field they convert to single photons, via the axion-photon-photon coupling in the Lagrangian $ \mathcal{L}_{a\gamma \gamma} = -g_{a\gamma \gamma}a\textbf{E}\cdot \textbf{B}$,
where $g_{a\gamma \gamma} =g_{\gamma} \alpha_\text{em} / \pi f_{a}$, $g_{\gamma}$ is a model dependent coupling coefficient, $\alpha_\text{em}$ is the electromagnetic fine-structure constant, $f_{a}$ is the axion decay constant, $a$ is the axion field, \textbf{E} is the electric field, and \textbf{B} is the magnetic field. 
Both major axion models, KSVZ and DFSZ, expect the coupling to be extremely weak, equal to $g_{\gamma}$ times 1/$f_a$, with $g_{\gamma} = 0.97$ and $-0.36$ respectively, and $f_a$ a very large scale. In the Sikivie haloscope a high quality factor microwave resonator is used to accumulate the axion to photon conversion signal \cite{Sikivie}. 
\begin{figure}[t]
\includegraphics[width=\linewidth]{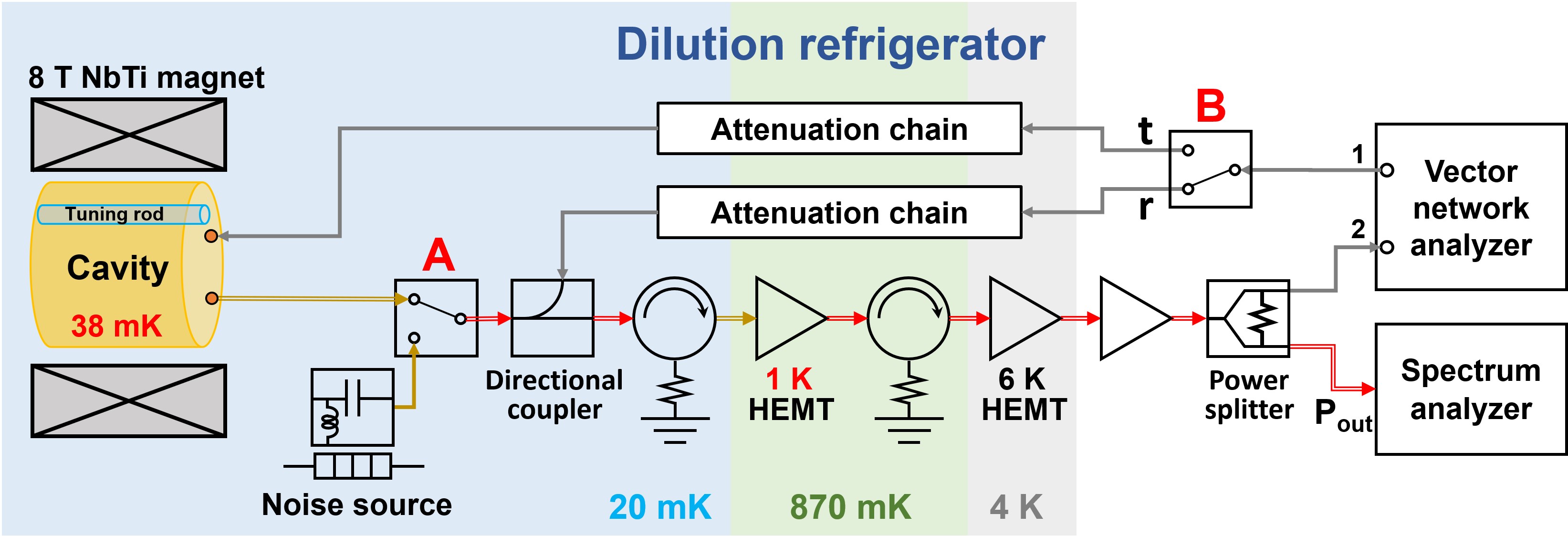}
\centering
\caption{ Schematic of the receiver chain of CAPP-PACE. The arrow lines indicate the rf transmission lines, the colored double lines the axion output path. Superconducting cables are marked with yellow lines. \textbf{A} switches the signals between the microwave cavity and noise source and \textbf{B} switches between the cavity transmission (\textbf{t}) and reflection (\textbf{r}) lines. \textbf{1} and \textbf{2} refer to the ports where the vector network analyzer (VNA) sends and receives signals, respectively. }
\label{fig:rfchain}
\end{figure}
In this case, the converted photons excite the resonant mode only if the photon frequency resides within the cavity bandwidth $\Delta\nu_c=\nu_c/Q_L$, where $Q_L$ is the loaded quality factor and $\nu_c$ is the cavity resonant frequency. %
The axion to photon conversion power, when the axion mass exactly matches $\nu_c$, is given by \cite{DKim_scanrate_2020}
\begin{equation}
\label{eq:pconv}
P_{a \rightarrow  \gamma \gamma} = \left(g^{2}_\gamma\frac{\alpha_\text{em}^{2}}{\pi^{2}}\frac{\rho_{a}}{\Lambda^{4}} \right) \left(\omega_{c} B^2 V C \frac{Q_0 Q_a}{Q_0 + Q_a}\right) , 
\end{equation} 
with the local dark matter density $\rho_{a} \cong 0.45$\,GeV/cm$^{3}$ \cite{Read_2014}, and $\Lambda \approx f_a^2 m_a^2\approx 78$\,MeV \cite{brubaker_thesis}.
The second parenthesis contains the following variable parameters: $\omega_c$ is the resonant angular frequency of the haloscope cavity, $B$ is the applied dc magnetic field, $V$ is the cavity volume, $C$ is the geometrical factor of the resonant cavity mode, $Q_0$ is the unloaded quality factor of the cavity, and $Q_a$ is the axion quality factor, estimated as $\sim 10^6$ \cite{prd_turner}.
The estimated axion signal power of one of the most powerful axion haloscopes so far is of the order of $10^{-23}-10^{-22}$\,W \cite{ADMX_PRL_DFSZ}. This makes a cryogenic low noise detecting system essential, in order to increase the signal to noise ratio (SNR).\\
\indent In a laboratory environment, the total noise temperature $T_{\text{sys}}$ of the axion haloscope detector is \cite{radiometer,noise_figure}
\begin{equation}
    \label{eq:PN}
    T_{\text{sys}} = T_{\text{phy}} + \frac{T_{\text{pre}}}{G_0} + \frac{T_{\text{else}}}{G_0 G_{\text{pre}}},
\end{equation}
where $T_\text{phy}=h\nu/k_\text{B}\left[{1}/{\left(e^{h\nu/k_\text{B}T_\text{cav}}-1 \right)}+1/2\right]$ is the noise from the physical temperature of the resonator (e.g, for the case $\nu=2.6$\,GHz and $T_\text{cav}=38$\,mK, $h\nu/2k_B=63$\,mK and $T_\text{phy}=67$\,mK),
$T_\text{pre}$ and $T_\text{else}$ represent the noise generated at the first amplifier and in the rest of the chain, respectively. $G_0$ is the reciprocal of overall attenuation before the preamplifier and $G_\text{pre}$ is the preamp gain. 
\\
%
%
%
\indent 
In the CAPP-PACE experiment, we focused to minimize $T_{\text{phy}}$, keeping in mind the quantum noise limited amplifier \cite{capp_jpa}. 
We used a Bluefors LD400 dilution refrigerator (DR) \footnote{Bluefors Oy, \url{https://bluefors.com}} which has a cooling power of 580\,$\mu$W when the mixing plate is at 100\,mK. 
We used a superconducting NbTi coaxial cable as the first transmission line after the antenna (Fig.\,\ref{fig:rfchain}) \footnote{The magnetic fringe field at the SC cable location is estimated to be less than 0.1\,T.}, acting as thermal insulation, blocking the heat flow between the plates at different temperatures. 
As a result, a cavity temperature of 38\,mK was achieved (measured using a calibrated RuO$_2$ thermometer \footnote{Lake Shore Cryotronics Inc.,\\ \url{https://www.lakeshore.com}}) even in an 8\,T magnetic field \footnote{American Magnetics Inc., \\ \url{http://www.americanmagnetics.com}}, i.e, for a 10.7\,$\mu$eV axion mass, $T_\text{phy}$ was $\sim10\%$ above the quantum limit.\\
\indent 
A HEMT amplifier with 40\,dB gain and 1\,K noise temperature \footnote{Low noise factory AB, \\ \url{https://www.lownoisefactory.com}} was used as shown in Fig.\,\ref{fig:rfchain}. 
The signal reduction before the preamplifier was minimized to an attenuation of $0.4\pm 0.1$\,dB. This effectively adds $\sim10\%$ more noise to the preamp noise ($1/G_0\simeq1.1$). 
The downstream noise contribution was estimated to be less than 1 mK [the third term in Eq.\,(\ref{eq:PN})].\\
\indent 
A high precision measurement system was built to determine the receiver chain noise. A 50\,ohm terminator was connected to switch \textbf{A} in Fig.\,\ref{fig:rfchain} serving as an accurate noise source. It was thermally linked to the PID-controlled heater, while being thermally separated from the mixing plate of the DR. Unlike the original \textit{Y}-factor method, which typically uses 2 points \cite{yfactor,noise_figure}, we used 795 temperature points in the interval between 200 and 960\,mK to mitigate the fitting error. While heating the noise source, the mixing plate temperature of the DR was maintained at $23\,(\pm\,2)$\,mK. We measured a total system noise temperature of 1$-$1.4\,K ($\pm$\,20\,mK) as shown in Fig. \ref{fig:noise_T}, dominated by the preamplifier. The observed noise increased at both ends of the frequency range because of the circulator operating range (2.1$-$2.6\,GHz) \footnote{Raditek Inc, \\ \url{https://raditek.com/raditek-circulator}}. \\
%
%
%
\indent
The haloscope cavity was designed to search axions in the frequency range from 2.457 to 2.749 GHz, i.e, the unexplored gap between existing results \cite{RBF1,RBF2}. 
\begin{figure}[t]
\centering
\includegraphics[width=\linewidth]{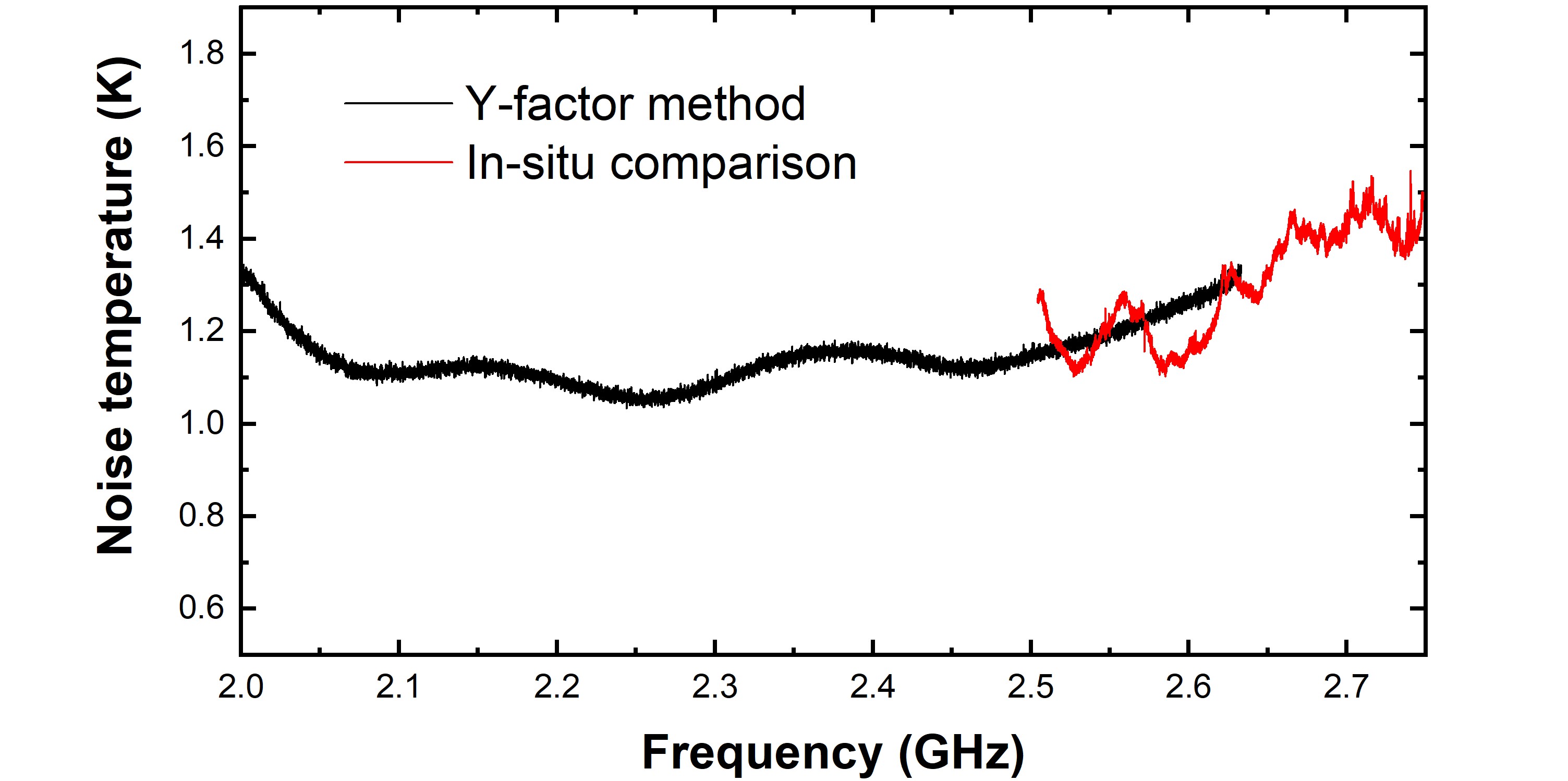}
\caption{Total system noise temperature versus frequency. The black line represents the system noise temperature obtained using the \textit{Y}-factor method utilizing the noise measurement setup shown in Fig.\,\ref{fig:rfchain}. The \textit{insitu} comparison method was used above 2.5\,GHz (red line) during the physics runs.}
\label{fig:noise_T}
\end{figure}
The cavity was composed of two half cavities each carved out of a solid cylindrical piece, shown in Fig.\,\ref{fig:split_cav}. 
This eliminates any electrical contact problems in the vertical direction, relevant to the TM$_\text{010}$ mode, the most commonly used cavity mode in axion haloscope searches due to its high geometrical factor.
The achieved quality factor was within $5\%$ of the maximum theoretical value allowed for copper. 
In addition, the eddy currents were significantly reduced by the electric insulation between the two halves, minimizing potential mechanical damage in case of a magnet quench. \\
\indent The resonant frequency of the cavity was tuned by moving the tuning rod from the center of the cavity toward the wall. To cover the whole range we applied 3 tuning rod configurations. 
First, we inserted a tuning rod made of low loss sapphire \cite{sapphire} to cover 2.457$-$2.500 GHz. 
When the sapphire rod was in the center $Q_0$ reached 130,000 in an 8\,T magnetic field. 
Second, we added a copper plated stainless steel rod in a fixed position and the resonant frequency was adjustable in the range of 2.500$-$2.605\,GHz. 
Third, the fixed rod was removed and the sapphire rod was replaced with copper. 
This allowed the resonant frequency to be tuned between 2.596 and 2.749\,GHz. Table\,\ref{tab:pace_spec} shows the parameters and experimental conditions employed for the tuning rod configurations. 
Above 2.5\,GHz the cavity is replaced with the larger volume cavity (see Fig.\,\ref{fig:split_cav}).
\\
\indent The resonant frequency was precisely adjustable with better than a kilohertz resolution by using Attocube piezo actuators \footnote{attocube systems AG, \url{https://www.attocube.com}}.
At most, 50\,$\mu$J of heat was generated by the piezo rotator, and the temperature rise of the cavity was less than 3\,mK at each tuning step. 
Within a few seconds, while the VNA was checking the resonant frequency, the temperature was restored to the value before tuning, so that the physical noise was kept stable near 40\,mK during the entire experimental runs. 
The direct contact between the lower end of the rotating shaft and the cavity bottom wall (Fig. \ref{fig:split_cav}) ensured fast temperature recovery and suppressed the hot-rod problem which can increase noise \cite{haystac_PRD}.\\
%
%
\indent 
The data reported in this letter were obtained from four independent runs, see Table\,\ref{tab:pace_spec}.
The experiments were divided into two approaches with one order of magnitude difference in sensitivity, and two orders of magnitude difference in frequency coverage. 
In the first approach, a wide range of 300\,MHz was scanned with $\sim 9\times \text{KSVZ}$ sensitivity (9KSVZ runs). 
The focus was on optimizing the routine of experimental processes such as cavity tuning, diagnosis, and data acquisition, to keep the cavity temperature as low as possible while minimizing dead time. 
In the second approach, a narrow range of 1\,MHz was scanned with close to the KSVZ sensitivity (KSVZ run). 
It confirmed the stability of the system by observing whether the noise of the data obtained for more than 10 h conformed to the Nyquist theorem \cite{sampling}.\\
\indent 
The physics data were collected by a commercial spectrum analyzer (SA) which is capable of analyzing up to 7\,GHz of high frequency microwave signals, and supporting a fast Fourier transform (FFT) mode \footnote{Rohde $\&$ Schwartz FSV 7 model, \url{https://www.rohde-schwarz.com}}. 
The actual data acquisition time versus elapsed time efficiency was slightly higher than $ 90\% $, but due to the windowing setting, described later, it took $89\%$ more time to obtain the same resolution bandwidth compared to no windowing \cite{windowing}. \\
\indent The data acquisition system (DAQ) software is described in Ref. \cite{daq_Lee_2017}.  
At the beginning of each run, it measures the system noise to act as reference for the \textit{insitu} noise calculation. 
At each frequency step it records the following parameters for calibration: the magnetic field $B$, cavity temperature $T_\text{cav}$, loaded quality factor $Q_L$, and coupling strength $\beta$ between the cavity and the receiver chain. 
The digitization of the averaged power spectrum coming out of the cavity is recorded next. 
In all runs, the bin width $\Delta\nu_b$ was set at 100 Hz so that an axion bandwidth $\Delta\nu_a$ contains more than 20 bins \cite{prd_turner}. 
The span was set in the range of 100$-$500\,kHz covering more than the cavity bandwidth. 
\\
\indent During DAQ, the real-time noise $T_\text{sys}^i$  was measured as an independent cross-check. 
In the $i$th tuning step, the total gain $G_\text{tot}^i$  of the receiver chain and the power amplitude $P_\text{off}^i$ of the off-resonant region of the data spectrum were measured, respectively. 
These values were compared with those in the reference step (zeroth step) and the real-time noise $T_\text{sys}^i$ was obtained as
\begin{equation}
\label{eq:T_realtime}
T_\text{sys}^i= T_\text{sys}^0 \frac{P_\text{off}^i}{P_\text{off}^0} \frac{G_\text{tot}^0}{G_\text{tot}^i}.
\end{equation} 
The red line in Fig.\,\ref{fig:noise_T} corresponds to the noise beyond the optimal working range of the circulator. 
The noise obtained with this method within the normal working range of the circulator is compatible with the noise previously measured using the \textit{Y}-factor method with a noise source. \\
\begin{table*}[ht]
\caption{\label{tab:pace_spec} Major parameters of the CAPP-PACE experiment. 
}
\begin{ruledtabular}
\centering
\begin{tabular}{ccccc}
Experimental run & 9KSVZ-1 & 9KSVZ-2 & 9KSVZ-3 & KSVZ \\ \hline
Period\footnotemark[1] (2018) & Jan 19$-$Feb 13  &   Jul 23$-$Aug 23   &  Nov 15$-$Dec 07 & Sep 01$-$Oct 26 \\
Frequency range & 2.457$-$2.500 GHz  & 2.500$-$2.605 GHz & 2.596$-$2.749 GHz & 2.5903$-$2.5918 GHz  \\
Mass ($m_a$) & 10.16$-$10.34\,$\mu$eV & 10.34$-$10.77\,$\mu$eV & 10.74$-$11.37\,$\mu$eV & 10.7126$-$10.7186\,$\mu$eV\\
Magnetic field ($B$) & 7.9\,T & 7.2\,T & 7.2\,T & 7.2\,T \\
Volume ($V$) &0.59\,L & 1.12\,L & 1.12\,L & 1.12\,L  \\
Tuning rod &sapphire rod & \begin{tabular}{@{}c@{}}sapphire rod \\[-3pt] +copper rod (fixed) \end{tabular} & copper rod & copper rod \\
Quality factor ($Q_0$) & 100\,\textit{k} & 80\,\textit{k} & 90\,\textit{k} & 100\,\textit{k}  \\
Geometrical factor ($C$)&0.62 & 0.51 &0.63 & 0.66   \\
Sweep time ($SWT$)\footnotemark[2]   & 18.92\,ms  & 18.92\,ms &  200\,ms\footnotemark[3] & 200\,ms\footnotemark[3] \\
Tuning step ($\Delta\nu$) & 16.0\,kHz  & 15.8\,kHz  & 16.2\,kHz  & 15.4\,kHz \\
Number of steps  & 2694 & 6642 &  9471 & 69\\
Number of spectra per step & 30\,\textit{k} & 10\,\textit{k} (5\,\textit{k}) & 15 (300) & 270\,\textit{k}\\
Sweep time per step ($\tau$)\footnotemark[4] & $\sim$10\,min  & 3\,min (1.5\,min) & 90\,s (60\,s) & 15\,h \\
\end{tabular}
\end{ruledtabular}
\footnotetext[1]{``Period" includes system management and upgrade time that occurred intermittently during the experiment.}
\footnotetext[2]{Defined as $W_\text{3dB}\times 1/\nu_b$. For the case of Blackmann-Harris windowing, $W_\text{3dB}=1.892$ \cite{windowing} and $SWT=W_\text{3dB}\times(1/100\,\text{Hz})=18.92$\,ms.}
\footnotetext[3]{If we set longer $SWT$ than the necessary time for the bin width ($=100$\,Hz), the SA applies 50$\%$ overlapping thus it gave a root-mean-squared spectrum from 21 [$\approx 200\,\text{ms}/ (18.92\,\text{ms}/2)$] overlapped spectra.}
\footnotetext[4]{$\tau=SWT\times$(Number of spectra per step).}
\end{table*}
%
%
\indent 
The basis of the analysis, the method and the procedure were similar to those of the ADMX, HAYSTAC, and CAPP microwave cavity axion experiments \cite{PRD_ADMX2001,haystac_PRL, capp8tb}. 
Most of the analysis process is devoted to constructing a grand spectrum normalized to the noise or target signal power over the entire scanned axion mass range.
There were a couple of subprocesses that had a significant impact on SNR. One was the fitting process, which eliminates non-uniform baselines resulting from various causes, such as a nonuniform gain distribution in a span and a slight impedance mismatch in the receiver chain, etc. 
In the 9KSVZ runs,
a five-parameter fit was used and the signal reduction was less than 5$\%$ with a stable Gaussian distribution \cite{PRD_ADMX2001}.   
In the KSVZ run, however, the 15 h of averaging exposed an unexpected morphology that the original parameter fitting could not describe. A well-developed Savitzky-Golay filter was used instead \cite{SGfilter} to obtain 85$\%$ SNR efficiency compared with the ideally distributed Gaussian noise. \\
\indent 
The other subprocess was to combine adjacent bins, called ``horizontal combination" \cite{haystac_PRD, capp8tb}. 
In this one, we cross-correlated the grand-spectrum with a virialized axion line shape \cite{prd_turner}. 
This process is mathematically the same as the maximum likelihood method in the previous experiment \cite{haystac_PRD}
\begin{figure*}[t]
\includegraphics[width=\linewidth]{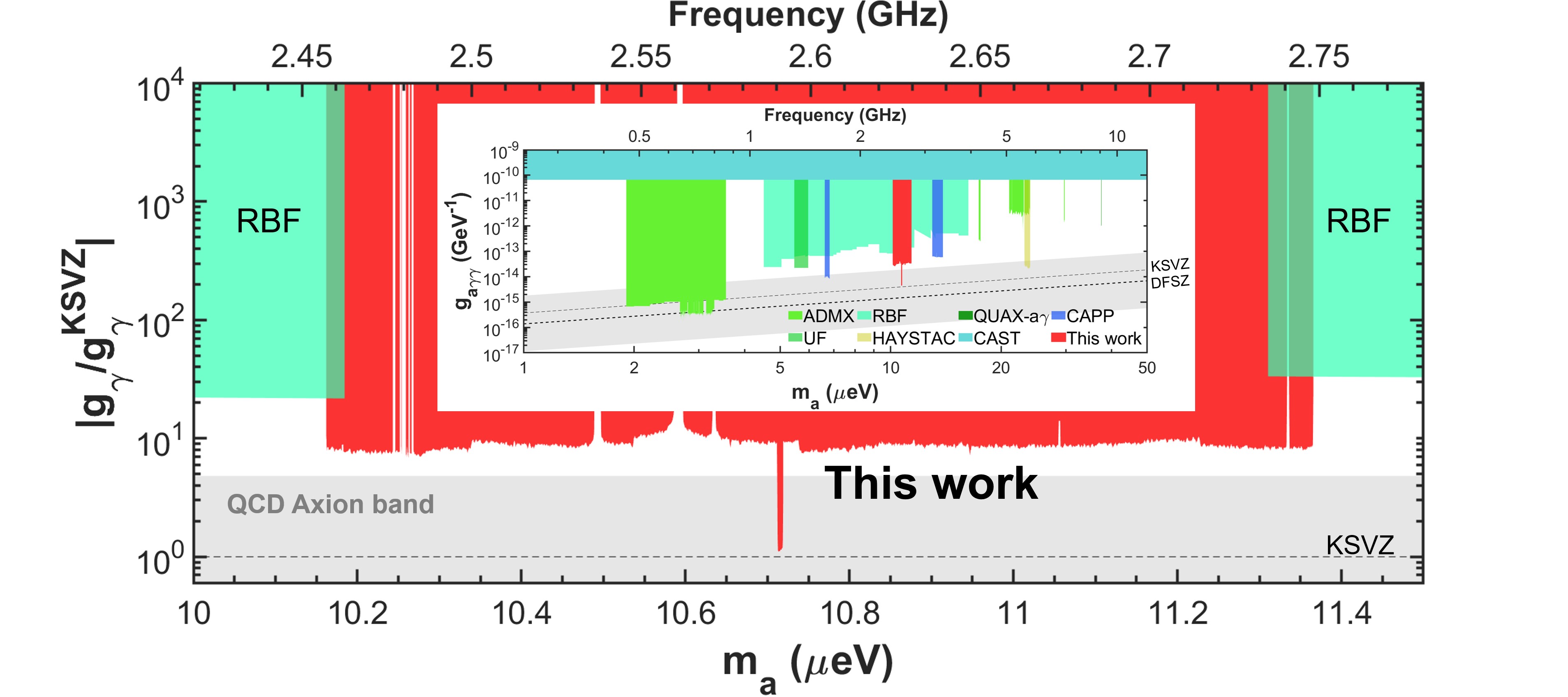}
\caption{The CAPP-PACE exclusion limit at $90\%$ confidence level (red area). Vacancy between the Rochester-Brookhaven-Fermilab (RBF) results (mint color gamut) \cite{RBF1,RBF2} is filled with this work. The inset shows this work along with other axion searching results in the extended axion mass range \cite{RBF1,UF,ADMX_PRL_1998,PRD_ADMX2001,PRD_ADMX2004,ADMX_PRL_DFSZ,ADMX_sidecar,haystac_PRL,cast_nature,Quax_agam,capp8tb,multicell_prl}.}
\label{fig:exclusion_plot}
\end{figure*}
but is differentiated due to the built-in function Blackman-Harris windowing of the SA \cite{windowing}.
Windowing gives larger weights in the middle of the single sweep time compared to the beginning and end, so that the effective bin size (value set in SA) is $W_\text{3dB}$ (3\,dB bandwidth of a window response function) multiplied by theoretical bin width $1/SWT$\,(sweep time). 
This creates correlations between nearby frequency bins in the Fourier transformed spectrum and significantly reduces the averaging efficiency when the virialized axion signal has a larger bandwidth than the bin size. 
The overlapping methods can partially offset the loss in efficiency \cite{windowing}. 
The SA uses a fixed overlapping ratio of 50$\%$, resulting in $\sim70 \%$ efficiency compared to ideal uniform windowed data (estimated using Monte Carlo simulation). The real-time spectrum analyzer will be used in the future. \\
\indent The SNR target was set at 5$\sigma$ with 90$\%$ confidence level for all runs. 
We had 81 candidates above $3.718\sigma$ for the 9KSVZ runs, and none for the KSVZ run. Each candidate was re-scanned for 30 min and none of them survived. 
Figure\,\ref{fig:exclusion_plot} shows the excluded axion mass range of 10.7126$-$10.7186\,$\mu$eV with close to KSVZ axion coupling sensitivity and 10.16$-$11.37\,$\mu$eV with 8$-$10 times KSVZ coupling at 90$\%$ confidence when we assumed a virialized axion line shape \cite{prd_turner}. The faintly visible gaps around 2.48\,GHz are due to bluetooth interference during the experiment. The gaps 
in 2.5$-$2.6\,GHz and near 2.74\,GHz, correspond to TE mode crossing. \\
%
%
%
\indent We reported the data establishing the first high sensitivity limits around 10.7\,$\mu$eV axions, which has never been previously explored. 
In this pilot experiment, an 8\,T superconducting magnet was used together with relatively small volume microwave cavities. 
Nevertheless, the axion scanning sensitivity was maintained at a high level by using a powerful dilution refrigerator, stable and high quality factor microwave cavities, a high resolution of frequency tuning system with low heat generation, and a low-noise HEMT amplifier. 
The cavity was successfully maintained near 40\,mK for all experimental runs, achieving the lowest physical temperature among all the axion experiments to date.\\
%
%
%
\begin{acknowledgments}
\indent This work was supported by IBS-R017-D1-2021-a00 of the Republic of Korea, and J.E.K was supported in part by the National Research Foundation (NRF) Grant No. NRF-2018R1A2A3074631. 
\end{acknowledgments}
%
%
%
\bibliography{reference}
\end{document}